\documentstyle[preprint,aps]{revtex}
\newcommand{\be}{\begin{equation}} \newcommand{\ee}{\end{equation}} 
\newcommand{\bea}{\begin{eqnarray}}\newcommand{\eea}{\end{eqnarray}}

\begin{document}
\draft
\preprint{MRI-PHY/01/96, hep-th/9601009}
\title { Exact Self-dual Soliton Solutions in a Gauged O(3) Sigma
Model with Anomalous Magnetic Moment Interaction}
\author{Pijush K. Ghosh$^{*}$}
\address{The Mehta Research Institute of
Mathematics \& Mathematical Physics,\\
Allahabad-211002, INDIA.}
\footnotetext {$\mbox{}^*$ E-mail: 
pijush@mri.ernet.in }  
\maketitle
\begin{abstract} 
It is shown that a gauged nonlinear $O(3)$ sigma model with
anomalous magnetic
moment interaction in $2+1$ dimensions is exactly integrable
for static, self-dual field configurations. 
The matter fields are exactly equivalent to those of the
usual ungauged nonlinear $O(3)$ sigma model. These static soliton
solutions can be mapped into an Abelian purely magnetic vortex solutions
through a suitable reduction of the non-Abelian gauge group.
A relativistic Abelian model in $2+1$ dimensions is also presented
where these purely magnetic vortices can be realized.
\end{abstract}
\narrowtext

\newpage

The nonlinear $O(3)$ sigma model in $2+1$ dimensions is popular
because of its applicability in various diverse branches of physics.
The soliton solutions of this model play a crucial role in most of
these applications. Fortunately, all the finite energy static, self-dual
soliton solutions of this model are well known which can be expressed in
terms of rational functions \cite{sig}. Interestingly enough, even when
the global symmetry is made local
to accommodate gauge field interaction in the theory and the gauge field
dynamics is solely governed by a non-Abelian Chern-Simons (CS) term, all the
static, self-dual soliton solutions can be written down explicitly
\cite{nardeli,kore}. In particular, the matter field solutions are same
for both the ungauged as well
as gauged nonlinear sigma model. These static soliton solutions
can be mapped into an Abelian purely magnetic vortex solutions through
a suitable reduction of the non-Abelian gauge group. These magnetic
vortices are exactly those found in a nonrelativistic CS theory \cite{jp}
except that the solitons in the nonrelativistic theory are charged also.
One can also identify these magnetic vortices as the
classical solutions of an Abelian theory in two dimensional Euclidean space
with a topological coupling \cite{nardeli}. However, no relativistic action
with dynamical gauge field in $2+1$
dimensions is known which reproduces these magnetic vortices.

In general, the gauge field dynamics in $2+1$ dimensions 
is governed by both Maxwell as well as CS term. The Maxwell
term can be dropped from the action only in the long
wave-length limit \cite{jatkar}. Thus, it is worth enquiring
at this point, whether or not the gauged nonlinear
$O(3)$ sigma model with
the gauge field dynamics governed by both the
Maxwell and the CS term, admits
static, self-dual soliton solutions.
It is known from the study of
vortex solutions in different Abelian as well
as non-Abelian Higgs model
that incorporation of anomalous magnetic moment interaction
\cite{paul,kogan} is necessary
in order to have self-dual soliton solution
in presence of both the Maxwell
and the CS term in the theory \cite{torres,piju}.

The purpose of this letter is to show
that the gauged sigma model with anomalous magnetic moment
interaction indeed
admits self-dual solitons. In particular, we find that the static, self-dual
soliton solutions of this model can be mapped to those of pure CS theory
considered in Ref. \cite{nardeli,kore}. Naturally, our solutions are
mapped into an Abelian purely magnetic vortex through a suitable reduction
of the gauge group. We also present a relativistic Abelian theory in
$2+1$ dimensions where these magnetic vortices can be realized.

Let us consider the following Lagrangian,
\be
{\cal{L}} = \frac{\lambda^2}{2} D_\mu \phi^a  D^\mu \phi^a -
\frac{1}{4} F_{\mu \nu}^a F^{a,\mu \nu} + \frac{\kappa}{4}
\epsilon^{\mu \nu \lambda} \left ( F_{\mu \nu}^a A_\lambda^a 
-\frac{e}{3}
\epsilon^{a b c} A_\mu^a A_\nu^b A_\lambda^c \right ) \ ,
\label{eq0}
\ee
\noindent where $a=1, 2, 3$ and the $\phi^a$'s are constrained to lie
on a unit two-sphere $S^2$, i.e. $\phi^a \phi^a = 1$.
We work here in Minkowskian space-time with the signature
$g_{\mu \nu}=( 1, -1, -1 )$ and the field-strength $F_{\mu \nu}^a$ is
$F_{\mu \nu}^a = \partial_\mu A_\nu^a - \partial_\nu A_\mu^a + e
\epsilon^{a b c} A_\mu^b A_\nu^c $. The velocity of light $c$ and the
Planck's constant in units of $\frac{1}{2 \pi}$ are taken to be unity.
The coefficient of the CS term ($\kappa$) and the coefficient of
the scalar kinetic energy term ($\lambda^2$) have mass dimension one.

The covariant derivative $D_\mu \phi^a$ is defined as
\be
D_\mu \phi^a = \partial_\mu \phi^a + \epsilon^{a b c} \left (
e A_\mu^b + \frac{g}{2} F_\mu^b \right) \phi^c ,
\label{eq1}
\ee
\noindent where the dual field-strength is given by,
\be
F_\mu^a = \frac{1}{2} \epsilon_{\mu \nu \lambda} F^{a, \nu \lambda} \ .
\label{eq2}
\ee
\noindent Note that there is an extra piece of the form
$\frac{g}{2} \epsilon^{a b c}  F_\mu^b \phi^c$ in the definition of the
covariant derivative. Such a non-minimal coupling for the gauge-field
is possible in $2+1$ dimensions which is also consistent with both 
Lorentz as well as gauge covariance of the theory \cite{paul,torres}.
The nonminimal coupling can be interpreted as the anomalous magnetic
moment interaction with $g$ identified as the anomalous magnetic
moment \cite{kogan}. In an Abelian theory, presence
of such a term in the covariant derivative generates a CS term after the
Higgs mechanism has taken place \cite{paul}. However, so far similar
mechanism has not been shown in the non-Abelian case.

The equations of motion which follow from (\ref{eq0}) are
\be
D_\mu J^{a,\mu} = 0,
\label{eq3}
\ee
\be 
\epsilon_{\mu \nu \lambda} \bigtriangledown^\mu \left (
F^{a,\lambda} + \frac{\lambda^2 g}{2 e} J^{a,\lambda} \right ) =
\lambda^2 J_\nu^a - \kappa F_\nu^a \ ,
\label{eq4}
\ee
\noindent where the current $J_\mu^a$ and the covariant derivative 
$\bigtriangledown_\mu \phi^a$ are defined as
\be
J_\mu^a = e \epsilon^{a b c} D_\mu \phi^b \phi^c \ , \ \
\bigtriangledown_\mu \phi^a = \partial_\mu \phi^a + 
e \epsilon^{a b c} A_\mu^b \phi^c \ . 
\label{eq5}
\ee
\noindent In obtaining
Eq. (\ref{eq3}) the constraint  $ \phi^a \phi^a = 1 $ is
taken care of by use of a Lagrange multiplier though not mentioned
explicitly in (\ref{eq0}).

Notice that the first order equation,
\be
\lambda^2 J_\nu^a = \kappa F_\nu^a \ ,
\label{eq6}
\ee
\noindent solves Eq. (\ref{eq4}) provided the following relation is
satisfied,
\be
g = - \frac{2 e}{\kappa} \ .
\label{eq7}
\ee
\noindent Eq. (\ref{eq6}) is identical to the corresponding equation
of \cite{nardeli}, except that now $J_\mu^a$ receives contribution from
the nonminimal part also. However, one can easily check using Eqs.
(\ref{eq5}) and (\ref{eq6}) that $J_\mu^a$ can also be rewritten as,
\be
J_\mu^a = \frac{{\tilde{J}}_\mu^a}{1 - \alpha^2} \ , \ \
{\tilde{J_\mu}^a} =
e \epsilon^{a b c} \bigtriangledown_\mu \phi^b \phi^c \ ,
\label{eq8}
\ee
\noindent where ${\tilde{J_\mu^a}}$ receives contribution
from the minimal part only and the dimensionless constant
$\alpha^2 = \frac{e^2 \lambda^2}{\kappa^2}$. Also note the identities,
\be
J_\mu^a \phi^a = 0 = F_\mu^a \phi^a \ ,\ \
( F_0^a )^2  = \frac{\lambda^2 \alpha^2}{(1 - \alpha^2)^2} 
( \bigtriangledown_0 \phi^a )^2 \ ,\ \
( F_i^a )^2  = \frac{\lambda^2 \alpha^2}{(1 - 
\alpha^2)^2} ( \bigtriangledown_i \phi^a )^2 \ ,  
\label{eq9}
\ee
\noindent which will be useful in our later discussions.

The energy functional $E$ can be obtained by varying (\ref{eq0}) with
respect to the background metric,
\be
E =  \frac{\lambda^2}{1 - \alpha^2} \int d^2 x \left [ \left
( \bigtriangledown_0 \phi^a \right )^2
+ \left ( \bigtriangledown_i \phi^a \right )^2 \right ] \ ,
\label{eq10}
\ee
\noindent where $F_{i 0}^a$ and $F_{12}^a$ have been eliminated by using
the identities (\ref{eq9}). It is interesting to note that
the energy functional (\ref{eq10}) is equivalent to the pure CS case
\cite{nardeli}
except for an overall multiplication factor. In order to ensure positivity
as well as finiteness of the energy functional, we choose
$0 < {\mid \alpha \mid} < 1$. The gauge fields completely decouple
from the theory at $\alpha=0$. In particular, one is left with either the
pure Yang-Mills CS (YMCS) action or the usual $O(3)$ sigma model plus
pure YMCS action with no interaction among the scalar and the gauge fields.
On the other hand, at $\alpha=1$, the Lagrangian (\ref{eq0})
admits zero energy trivial vacuum solutions.

The energy functional (\ref{eq10}) can be rearranged as \cite{bogo}, 
\be
E = \frac{\lambda^2}{1-\alpha^2} \int d^2x \left [ \left
( \bigtriangledown_0 \phi^a \right)^2 + \frac{1}{2} \left (
\bigtriangledown_i \phi^a \mp \epsilon_{ij} \epsilon^{a b c}
\phi^b \bigtriangledown_j \phi^c \right )^2 \right ] \\ +
\frac{8 \pi \lambda^2}{1 - \alpha^2} {\mid Q \mid} \ ,
\label{eq11}
\ee
\noindent where $Q = \int d^2 x K_0 $ is the topological charge and $K_0$
is the zeroeth component of the topological current $K_\mu$,
\be
K_\mu = \frac{1}{8 \pi} \int d^2 x \epsilon_{\mu \nu \lambda} 
\epsilon^{a b c} \phi^a \partial_\nu \phi^b \partial_\lambda \phi^c \ .
\label{eq12}
\ee
\noindent The energy in Eq. (\ref{eq11}) has a lower bound
$E \geq \frac{8 \pi \lambda^2}{1 - \alpha^2} {\mid Q \mid}$ in terms of the
topological charge.
The bound is saturated when the following Bogomol'nyi equations \cite{bogo}
are satisfied,
\be
\bigtriangledown_0 \phi^a = 0, \ \
\bigtriangledown_i \phi^a \mp \epsilon_{ij} \epsilon^{a b c} \phi^b
\bigtriangledown_j \phi^c = 0 \ .
\label{eq13}
\ee
\noindent These Bogomol'nyi equations are
consistent with the second order field equations (\ref{eq3}).

The first equation of (\ref{eq13}) implies that the Noether charge is
zero.  It also follows from Eqs. (\ref{eq6})
and (\ref{eq13})
that for static soliton solutions $A_0^a$ must have the form,
\be
A_0^a = \frac{\kappa e}{\lambda^2 ( 1- \alpha^2)} \phi^a \ .
\label{eq14}
\ee
\noindent Note that $A_0$ has the same expression as in the case of pure
CS theory \cite{nardeli} except an overall multiplication factor.
As a consequence of vanishing Noether charge, the zeroeth component of
the gauge field equation (\ref{eq6}) dictates that the spatial
component of the gauge fields are pure gauges,
\be
A_i^a = i Tr \left ( \sigma^a U^{- 1} \partial_i U \right ),
\label{eq15}
\ee
\noindent where gauge group element $U= exp[-i \eta^a(x) \sigma^a/2]$,
$\sigma^a$ being the Pauli matrices. The second equation of (\ref{eq13})
can now be rewritten in terms of the $U$-gauge transformed field
$\phi \rightarrow U^{-1} \phi \ U$ as,
\be
\partial_i \phi^a \mp \epsilon_{ij} \epsilon^{a b c} \phi^b
\partial_j \phi^c = 0 \ ,
\label{eq16}
\ee
\noindent which is nothing but the self-dual equation for the ungauged
nonlinear $O(3)$ sigma model.
Using the stereographic projections,
\be
u_1 = \frac{\phi_1}{1+\phi_3}, \ \ u_2 = \frac{\phi_2}{1+\phi_3} \ ,
\label{eq17}
\ee
\noindent where $u = u_1 + i u_2$ is a complex-valued function, Eq.
(\ref{eq16}) can be conveniently written as,
\be
\partial_i  u = \pm i \epsilon_{ij} \partial_j  u \ .
\label{eq18}
\ee
\noindent Any analytic (anti-analytic) function $u(z)$ is a solution of
the equation
(\ref{eq18}). So, the minimum-energy matter field solutions of the
Lagrangian (\ref{eq0}) are exactly equivalent to those of the gauged
sigma model with a CS term \cite{nardeli}
and, hence to the ungauged sigma model \cite{sig}.
The only effect due to the
incorporation of anomalous magnetic moment interaction
is the change in energy
by an overall multiplication factor.

It is known that these non-Abelian solutions give rise to nontrivial vortex
configuration under a suitable contraction of the gauge group \cite{nardeli}.
In particular, one can construct a gauge invariant Abelian
field strength \cite{'thooft} as,
\bea
{\cal{F}}_{\mu \nu} \ & = & \ \frac{1}{e}
\epsilon^{a b c} D_\mu \phi^a D_\nu \phi^b
\phi^c - \phi^a F_{\mu \nu}^a\nonumber \\
& = & \ \frac{1}{e (1-\alpha^2)^2} \epsilon^{a b c}
\bigtriangledown_\mu \phi^a \bigtriangledown_\nu \phi^b
\phi^c - \phi^a F_{\mu \nu}^a \ .
\label{eq20.1}
\eea
\noindent The last term of (\ref{eq20.1}) vanishes identically because of
the first identity of (\ref{eq9}). Now note that ${\cal{F}}_{i 0}=0$, since
$D_0 \phi^a = 0$. However, the magnetic field $B=-{\cal{F}}_{12}$ is nonzero,
\be
B= \pm \frac{4}{e (1-\alpha^2)^2} \frac{{\mid u^\prime(z) \mid}^2}{
\left ( 1 + {\mid u(z) \mid}^2  \right )^2 } \ ,
\label{eq20.2}
\ee
\noindent where $u^\prime(z)= \frac{\partial u}{\partial z}$. The magnetic
flux $\Phi=\int d^2 x B= \pm \frac{4 \pi}{e(1-\alpha^2)^2} {\mid Q \mid}$
is quantized in terms of the topological charge $Q$.
Note that $B$ solves the Liouville equation. This is also the expression
for the magnetic field in $2+1$ dimensional nonrelativistic gauged
nonlinear Schr$\ddot{o}$dinger equation \cite{jp} which describes
magnetic, charged vortices. The reproduction of the magnetic vortices
(\ref{eq20.2})
in an Abelian theory is not that surprising as the field strength
${\cal{F}}_{\mu \nu}$ satisfies the Bianchi identity $\epsilon_{\mu
\nu \lambda} \partial^\mu {\cal{F}}^{\nu \lambda}=0$ in $2+1$ dimensions.

At this point one would like to look for relativistic Abelian theories
which reproduce the magnetic vortices (\ref{eq20.2}). Though a
Euclidean two dimensional model reproducing the magnetic vortices
(\ref{eq20.2}) is known \cite{nardeli}, no $2+1$ dimensional
relativistic Abelian model with dynamical gauge field is known
to give these magnetic vortices. Here, we present one such
$2+1$ dimensional relativistic Abelian model.

Consider the Lagrangian
\be
{\cal{L}} = - \frac{G(\chi^3)}{4} F_{\mu \nu} F^{\mu \nu}
+ \frac{\lambda^2}{2} \left ( {\cal{D}}_\mu \chi^a \right )^2
- \frac{\lambda^4 e^2}{2 G(\chi^3) } \left ( 1 + \chi^3 \right )^2 \ ,
\label{eq21}
\ee
\noindent where $\chi^a$ 's are constrained to lie on the unit two-sphere
$S^2$, i.e. $\chi^a \chi^a = 1$. The covariant derivative is defined as,
\be
{\cal{D}}_\mu \chi^a = \partial_\mu \chi^a +
e \epsilon^{a b c} A_\mu n^b \chi^c \ ,
\label{eq22}
\ee
\noindent where $n^a$ 's are the components of a constant unit vector
$n=(n^1, n^2, n^3)=(0, 0, 1)$. The Lagrangian (\ref{eq21}) represents
a nonlinear $O(3)$ sigma model with its $U(1)$ subgroup gauged where the
gauge field dynamics is governed by a Maxwell term \cite{dur,gg,ua,gg1}. The
Lagrangian (\ref{eq21}) exactly reduces to the one considered in \cite{dur}
when $G(\chi^3)=1$ and $\chi^a$ is replaced by $- \chi^a$. The
modification to the Maxwell kinetic term in (\ref{eq21}) can be viewed as
an effective action for a system in a medium described by a suitable
dielectric function $G(\chi^3)$. In fact, such a dielectric function is used
in connection with the soliton bag models of quarks and gluons \cite{lee}.
Incorporation of such a nonminimal interaction in the action is also
necessary, in certain supersymmetric gauge theories with non-compact gauge
group, in order to have a sensible gauge theory \cite{hull}. This nonminimal
coupling is interesting in the context of vortex solutions also as one can
have infinitely degenerate topological vortex solutions where the magnetic
flux is not necessarily quantized \cite{piju}.

The Lagrangian (\ref{eq21}) admits self-dual solutions for arbitrary
$G(\chi^3)$. However, here we restrict ourselves to a particular form
of $G(\chi^3)$, namely,
\be
G(\chi^3) = \frac{(1+\chi^3)^2}{1-\chi^3} \ ,
\label{eq22.1}
\ee
\noindent for which the scalar potential ( the last term of (\ref{eq21}) )
becomes,
\be
V(\chi^3) = \frac{e^2 \lambda^4}{2} \left ( 1 - \chi^3 \right ) \ .
\label{eq22.2}
\ee
\noindent Note that this potential is usually used in
describing Skyrmions \cite{sky}. The finite energy field configurations
demand that $\chi^3$ should take the vacuum value,
i.e. $\chi^3 = 1$, at spatial infinity. The scalar fields $\chi^1$
and $\chi^2$ have the same mass $m=\frac{e \lambda}{\sqrt{2}} $,
where as $\chi^3$ is massless. Note that the gauge field $A_\mu$
is also massless as the local $U(1)$ symmetry is not
broken spontaneously due to the constraint $\chi^a \chi^a=1$.
The dielectric function (\ref{eq22.1}) is singular at $\chi^3 = 1$.
However, the relevant term in the energy functional, i.e.
$G(\chi^3) F_{12}^2$, is non-singular at $\chi^3=1$ for
the specific solutions we obtain below.

The static energy functional corresponding to the Lagrangian (\ref{eq21}) in
the gauge $A_0=0$ can be written as,
\bea
E \ & = & \ \frac{1}{2} \int d^2 x \left [  \lambda^2 \left
( {\cal{D}}_i \chi^a \pm
\epsilon_{ij} \epsilon^{a b c} \chi^b {\cal{D}}_j \chi^c \right )^2
+ G(\chi^3) \left ( F_{12} \pm e \lambda^2 \frac{1 -
\chi^3}{1+\chi^3} \right )^2
\right ]\nonumber \\
& & \pm 4 \pi \lambda^2 \int d^2 x k_0,
\label{eq23}
\eea
\noindent where $k_0$ is the zeroeth component of the topological
current,
\be
k_\mu = \frac{1}{8 \pi} \epsilon_{\mu \nu \lambda} \left [
\epsilon^{a b c} \chi^a {\cal{D}}^\nu \chi^b {\cal{D}}^\lambda \chi^c
- e F^{\nu \lambda} ( 1 + \chi^3) \right ] \ .
\label{eq24}
\ee
\noindent The energy functional (\ref{eq23}) has a lower bound in terms
of the topological charge $q=\int d^2 x k_0$, $E \geq 4 \pi q \lambda^2$.
The bound is saturated when the following first order equations hold true,
\be
{\cal{D}}_i \chi^a \pm \epsilon_{ij} \epsilon^{a b c}
\chi^b {\cal{D}}_j \chi^c = 0, \ \
F_{12} \pm e \lambda^2 \frac{1 - \chi^3}{1+\chi^3} = 0 \ .
\label{eq25}
\ee
\noindent Using the stereographic projections (replace $\phi^a$
by $\chi^a$ in (\ref{eq17}) and identify $u(z)$ as $\Omega(z)$)
one can see that the two equations in
(\ref{eq25}) reduces to the Liouville equation,
\be
\bigtriangledown^{2} ln {\mid \Omega \mid}^2 = - e^2 \lambda^2
{\mid \Omega \mid}^2 \ ,
\label{eq26}
\ee
\noindent all of whose solutions are well known. In particular,
\be
{\mid \Omega \mid}^2 = \frac{8}{e^2 \lambda^2}
\frac{{\mid w^\prime(z) \mid}^2}{
\left ( 1 + {\mid w(z) \mid}^2  \right )^2 } \ ,
\label{eq26.1}
\ee
\noindent where $w(z)$ is an arbitrary analytic (anti-analytic) function and
$w^\prime(z)=\frac{\partial w(z)}{\partial z}$. 

Note from the second equation of (\ref{eq25}) that ${\cal{B}}= -F_{12}
= \pm e \lambda^2 {\mid \Omega \mid}^2$. Thus, the magnetic field $B$ can
be realized by the static, self-dual soliton solutions of (\ref{eq21}).
The most general $N$-soliton
solutions (\ref{eq26.1}) are described by $4 N$ real parameters
\cite{jp,kor}. The radially symmetric solutions can be obtained
by putting $w(z)= c_0 z^N, z=r e^{i \theta}$ in (\ref{eq26.1}), where
$c_0$ is an arbitrary constant related to the size of the solitons.
It is worth mentioning at this point, though the soliton solutions of the
non-Abelian theory are topological in nature, the soliton solutions
of (\ref{eq21}) are not. This can be seen as follows. The topological
current $k_\mu$ can be written as ( replace $\phi^a$ by $\chi^a$ in
$K_\mu$ ),
\be
k_\mu = K_\mu -
\frac{e}{4 \pi} \epsilon_{\mu \nu \lambda} \partial^\nu
\left [ A^\lambda ( 1 + \chi^3 ) \right ] .
\label{eq27}
\ee
\noindent The first term in (\ref{eq27}) defines the nontrivial mapping
$S^2 \rightarrow S^2$. As a result, the stability of the soliton solutions
are guaranteed when the topological
charge $q=\int k_0 d^2 x$ receives contribution from the first term of
(\ref{eq27}) only. On the other hand, if the contribution to $q$ is from
the second term only, the solitons are nontopological (i.e. no topological
arguments can be made for their stability ). Since, $\chi^3 \rightarrow 1$
at both the spatial infinity for the soliton solutions
of (\ref{eq21}), only the second term of (\ref{eq27}) contributes
to $q$. As a result, the soliton solutions (\ref{eq26.1}) are of
nontopological in nature.

Finally, the following comments are in order,

(i) The scalar potential $V(\chi^3)$ is uniquely determined for
self-dual soliton solutions, once a particular $G(\chi^3)$ is chosen.
Consequently, for different choices of $G(\chi^3)$, the decoupled
equation (\ref{eq26}) can be reduced to well known integrable equations
in two Euclidean space and, hence the Lagrangian (\ref{eq21}) admits exact
solutions for a class
of dielectric function. It is worth mentioning at this point that
similar abelian Higgs models with a dielectric function can also be
constructed  which are integrable for static, self-dual field
configurations \cite{piju1}.

(ii) The partially gauged nonlinear $O(3)$ sigma model of the
type (\ref{eq21}) was considered in the literature with the prior
motivation of
breaking the scale invariance of its nongauged counterpart \cite{dur,gg}.
However, notice that the soliton solution of (\ref{eq21}) as given
in (\ref{eq26.1})
is indeed scale invariant. Thus the scale invariance of soliton solutions
in different gauged sigma models are model dependent.

The Lagrangian (\ref{eq21}) can well be considered in the two Euclidean
space since we are looking for static soliton solutions in the gauge
$A_0=0$. Naturally, the self-dual equations
of the Euclidean action in two dimensions corresponding to (\ref{eq21})
are given by (\ref{eq25}). We would like to point out that in this case
the trace of the energy-momentum tensor of our Euclidean action vanishes
at the self-dual limit.

(iii) The Lagrangian (\ref{eq0}) can be equivalently formulated
as a gauged $CP^1$ model using the Hopf map $\phi^a=Z^\dagger
\sigma^a Z$, where $Z$ is a two component complex scalar field.
Now generalizing this gauged $CP^1$ model to the
$CP^N$ case, we have,
\be
{\cal{L}} = \frac{\lambda^2}{2} \left ( {\mid D_\mu Z \mid}^2 -
{\mid Z^\dagger D_\mu Z \mid}^2 \right ) 
- \frac{1}{4} F_{\mu \nu}^a F^{a,\mu \nu} + \frac{\kappa}{4}
\epsilon^{\mu \nu \lambda} \left ( F_{\mu \nu}^a A_\lambda^a
-\frac{e}{3} f^{a b c} A_\mu^a A_\nu^b A_\lambda^c \right ) \ ,
\label{eq28}
\ee
\noindent where $Z$ is a $(N+1)$-component complex scalar field
with the constraint $Z^\dagger Z = 1$. The
covariant derivative in (\ref{eq28}) is defined as $D_\mu Z =
( \partial_\mu - i ( e A_\mu^a + \frac{g}{2} F_\mu^a ) \Gamma^a ) Z$,
where $A_\mu^a$'s are the $SU(N+1)$ gauge fields and the
$\Gamma^a$'s, the generators of the $SU(N+1)$, are the traceless
hermitian matrices. The Lagrangian (\ref{eq28}) admits static, self-dual
soliton solutions for arbitrary $N$ at $g=-\frac{2 e}{\kappa}$. The complex
scalar field $Z$ is exactly equivalent to that of the usual ungauged
$CP^N$ model at the self-dual point. The spatial components of the gauge
fields are pure gauges for these self-dual solitons.
However, $A_0^a$ is determined as $A_0^a = \frac{\kappa e}{\lambda^2
(1-\alpha^2)} Z^\dagger \Gamma^a Z$. This shows that the static,
self-dual soliton solutions of the gauged $CP^N$ model (\ref{eq28})
can be mapped into the corresponding solutions of pure CS theory
\cite{nardeli,kore}. One would expect that the soliton solutions of
these two models are quite different in nature, away from
the self-dual point. It would be interesting to study the soliton
solutions of these two models away from this point.

\acknowledgements

I thank Dileep Jatkar, Avinash Khare and Sumathi Rao for a careful
reading of the manuscript and valuable comments.

\end{document}